\documentclass[twocolumn]{aastex631}

    \usepackage{tabularray}
    \usepackage{spverbatim}

\usepackage{academicons}
\definecolor{orcidlogocol}{HTML}{A6CE39}
\usepackage{hyperref}
\usepackage{orcidlink}

\begin{document}

\title{Searching for Dwarf Novae in 47 Tucanae}

\author{Iman Behbehani\orcidlink{0009-0008-1688-2323}}
\affiliation{The Graduate Center at the City University of New York,365 5th Ave, New York, NY 10016}
\affiliation{Center for Computational Astrophysics, Flatiron Institute, 62 5th Ave 9th floor, New York, NY 10010}
\affiliation{Department of Astrophysics, American Museum of Natural History,200 Central Park W, New York, NY 10024}

% %\collaboration{20}{(705 Thesis Paper)}

\author{M. Shara\orcidlink{0000-0003-0155-2539}}
\affiliation{Department of Astrophysics, American Museum of Natural History,200 Central Park W, New York, NY 10024}

\author{D. Zurek\orcidlink{0009-0008-4599-2935}}
\affiliation{Department of Astrophysics, American Museum of Natural History,200 Central Park W, New York, NY 10024}

\begin{abstract}

The globular cluster 47 Tucanae has 43 strong cataclysmic variable candidates, four of which are confirmed dwarf novae. Tidal capture theory predicts the existence of dozens of dwarf novae, suggesting more are undiscovered. Using $\sim 400$ nm long exposure images from \textit{Hubble}’s ACS camera, we searched five epochs of archival \textit{Hubble} images and did not find any variability at the scale of a dwarf nova eruption. We found 1 previously undiscovered variable object on the main sequence that could not be classified as a CV; the object experienced a change of $\approx$ 0.3 Mag in 425 nm, and $\approx$ 0.2 mag in 635 nm over the course of 12 days. The lack of new CV detections in 47 Tucanae could be a result of the short timescale of the DN eruption, and a low observing probability. However, this null result may also support research suggesting most CVs in globular clusters are not dynamically formed, and tidal capture theory overestimates the detectable CV population.

\end{abstract}



\section{Introduction} \label{sec:intro}
Stars are drawn to each other gravitationally, and form both large and small stellar systems. Globular Clusters (GCs) are densely packed, gravitationally bound groups of tens of thousands to millions of stars. Most galactic GCs are around 13 Gyr old \citep{Vandenberg}. GCs are considered crowded fields due to the spatial density of stars. This proximity often leads to the formation of binaries (see, e.g., \citet{Hut;1992}). Close binaries in GCs can impact the evolution of a GC through collisions, mass transfer, mergers, and tidal capture \citep{Stodolkiewicz;1986}.

Cataclysmic variable stars (CVs) are interacting binary star systems composed of a white dwarf (WD) and a main sequence (MS) companion star. These companions become attached to the white dwarf via primordial binary stellar evolution \citep{robinson}, or two- or three-body tidal capture or exchange \citep{Fabian,Michaely}. These interactions are most likely to occur in dense stellar environments like globular clusters. However, recent research questions the role of dynamic formation of CVs in GCs \citep{Belloni;2016,Belloni2020,RiveraSandoval;2024}, and suggests this overestimates the presence of bright CVs in GC. 

The magnetic field strength of the WD impacts the way mass is transferred from the MS star. Magnetic CVs generally accrete in the direction of their magnetic field lines. Non-magnetic (or weak magnetic) WDs will instead form an accretion disc. In a subgroup of non-magnetic CVs, thermal instabilities in this disk cause periodic outbursts called dwarf novae (DN) (see, e.g., \citet{Lasota}), characterized by a 2-20 day eruption with an amplitude around 2-5 magnitudes.

47 Tucanae (47 Tuc) is one of the most massive of the ~150 GCs in the Milky Way \citep{Harris, Baumgardt2018}. It has become one of the most frequently observed GCs in the Milky Way, accumulating a wealth of accessible Hubble Space Telescope data. 47 Tuc, also known as NGC 104, contains at least 500,000 stars, of which 43 pairs are CV candidates (on the basis of their X-ray, ultraviolet and optical properties) \citep{RiveraSandoval, RiveraSandoval;2024}.

% GCs are mostly populated by old, red main sequence stars, all of which are less luminous than the Sun. However, the luminosities of GCs are dominated by the $\approx1$ percent of stars that are red giants. Red giants have high luminosities  (100 to 1000 $L_\odot$) which make them extremely prominent in images of GCs. The glare of red giants makes searching for dwarf nova eruptions difficult in red optical filter images of GCs. The outbursts of dwarf novae are typically dominated by blue light, so we have focused on shorter wavelength \textit{Hubble Space Telescope (HST)} images in our search for new dwarf novae in eruption.

Dwarf novae have been detected in GCs during eruptions with time domain observations \citep{Knigge;2003,Shara;2005}. \citet{sharaberg} detected 1 candidate DN over 12 near-UV \textit{HST} epochs of 47 Tuc. \citet{Servillat;2011} found 1 DN in 5 epochs of observations lasting one to a few days. The observed rarity of DN in these studies may be tied to the small fraction of the DN cycle where the DN is in outburst. Other literature has suggested detecting DN in quiescence (non-eruption) \citep{Belloni;2016} via $H\alpha$ imaging \citep{Cool;1995,Grindlay;1995,Pietrukowicz;2009}, or far-ultraviolet (FUV) detection \citep{Dieball;2010}. The accretion process makes the CV system emit at shorter wavelengths than other stars typically in GCs. FUV imaging also will reduce detection of main sequence stars in the crowded field emitting in the infrared.

47 Tuc is estimated to have $\sim$100 CVs solely from two-body tidal capture \citep{distef}. This early estimate excluded CVs formed by three-body tidal capture, so the estimated number in fact exceeds 100. However, only 43 identified detectable CV candidates are currently known in 47 Tuc, with only 4 observed erupting as dwarf novae \citep{RiveraSandoval,RiveraSandoval;2024}. This study attempts to detect more CVs as erupting DN through unsearched optical archival HST data around 400nm. A non detection could indicate a lack of remaining CVs for detection, support alternative methods for detection methodology, or suggest CVs are not dynamically formed in globular clusters, and the estimate from tidal capture theory is too high.

\section{Searching for Variables} \label{sec:style}
\subsection{Image Selection}
 Key identifiers of erupting dwarf novae are large increases in brightness (typically 10 to 100 times) for an interval of 3 to 15 days. We searched for stars with large brightness increases by visually blinking \textit{HST} epochs with identical filters and similar exposure times. Images were aligned in the \textit{FITS} file viewing software \textit{SAOImage DS9}, and aligned using WCS coordinates. The coordinates of all stars exhibiting changes in brightness were recorded as potential variables, and compared with those of previously known CV candidates. We adjusted image scale in \textit{DS9} to account for exposure differences, and conducted a photometric analysis (on unaltered data) on sources suspected to be variable.
 
 We used short wavelength observations $\sim$400nm to reduce crowding from redder MS stars, and highlight bluer WDs in CVs. Longer exposure images include fainter stars, and were preferred to make our searches more sensitive to faint CVs. All image pairs were taken at least two weeks apart to allow for potential dwarf nova eruptions to fully begin or subside by the subsequent image. These criteria left us with 5 usable images from the \textit{Hubble} \textit{Advanced Camera for Surveys} (\textit{ACS}) camera. The images were selected from Hubble's \textit{MAST}'s public online database, and are listed in Table \ref{table:epochs} with their filter, date, and exposure times. Images were taken between 2002 and 2012, with exposure times from 150s to 1900s. We used  calibrated and charge transfer efficiency (CTE) corrected \texttt{drc.fits} files in this study, minimizing the impact of solar particles and cosmic rays in our data.

\begin{deluxetable}{ccccc}\label{table:epochs}
\tablecaption{Epochs used in this study. Datasets J8FW01010, J8FW2010, J8FW03010, J8CD01010, and JBXG03020 were searched for variables. Data taken at F625W were incorporated for further photometric analysis. J6LL01010 and J6LL01YGQ were used to make a color magnitude diagram of the cluster. \label{epochs}}
\tablehead{
\colhead{Image} & \colhead{Camera} & \colhead{Date} & \colhead{Exp (s)} & \colhead{Filter}
}
\startdata
% J6LL01YIQ & ACS & 2002-07-07 & 150  & \textsc{F435W} \\
J8FW01010 & ACS & 2002-09-30 & 300  & \textsc{F435W} \\
J8FW2010 & ACS & 2002-10-02 & 310  & \textsc{F435W} \\
J8FW03010 & ACS & 2002-10-11 & 345  & \textsc{F435W} \\
J8CD01010 & ACS & 2002-04-05 & 900  & \textsc{F475W} \\
J6LL01010 & ACS & 2002-07-07 & 450 & \textsc{F475W} \\
JBXG03020 & ACS & 2012-07-29 & 1900 & \textsc{F475W} \\
J6LL01YGQ & ACS & 2002-07-07 & 150 & \textsc{F550W} \\
J8FW01020 & ACS & 2002-9-30 & 465 & \textsc{F625W} \\
J8FW02020 & ACS & 2002-10-02 & 455 & \textsc{F625W} \\
J8FW03020 & ACS & 2002-10-11 & 400 & \textsc{F625W} \\
\enddata
\end{deluxetable}

% Based on a search of the available \textit{ACS} database, our exploration has expanded the sample size from the original study from \citet{RiveraSandoval} and \citet{sharaberg} with the unsearched data, expanding and detailing the prior research. Thus, we are looking to see if any new detectable dwarf novae exist in the unexplored data.

\vspace{5mm}

\subsection{Image Searching}
Images were paired by similarities in exposure and wavelength and compared both visually and computationally. After the images were normalized and aligned, we differenced them to identify any variability. We conducted an additional visual investigation in which images were aligned in \textit{SOAImage ds9} according their \textit{WCS} data and normalized. Any potential variables were inspected directly.

We conducted a photometric analysis for stars that appeared to be variable. Epochs were aligned using \texttt{Astrodrizzle} \citep{Hack;2012}. Potentially variable sources were studied with \texttt{PhotutilsAperture} packages for \texttt{CircularAperture}, \texttt{AperturePhotometry}, and \texttt{ApertureStats} functions \citep{bradley_2026_19636730}. Intensities within a circular radius of 3 pixels from the center of the star were summed and compared between epochs to monitor variability. We used relative magnitudes to ensure the detected variability was not a result of observational differences between epochs. We use the following formula to convert from image fluxes to relative magnitudes, where $m_1$ and $m_2$ represent apparent magnitudes of the variable source and the background star, and $F_1$ and $F_2$ represent their fluxes \citep{Carroll;1996}.

\begin{equation}
    m_1 - m_2 = -2.5 \log(\frac{F_1}{F_2})
\end{equation}

A star's change in relative magnitude between epochs may indicate a dwarf nova or other stellar eruption. We calculate the relative magnitudes of the variable (referred to as `source') and two selected background stars (referred to as S1 and S2). If variability is detected, S1-Source and S2-Source will change between epochs while S1-S2 is constant.

\section{Results}
After searching all five epochs, no new DN were found. One unknown variable source was detected between J8FW01010 and J8FW03010, but it does not meet the necessary change in magnitude to be a dwarf nova. 

Observations of 47 Tucanae at 625nm were serendipitously available in the \textit{HST} archive at the same dates as the F435W observations. We use this additional set of epochs for photometric analysis at the same dates, 2002-09-30, 2002-10-02 and 2002-10-11, in a different filter. These epochs were also taken with the \textit{HST} \textit{ACS} instrument, and were observed with exposures 465, 455, and 400 seconds (see Table \ref{table:epochs}).

\begin{figure}[ht!]
\plotone{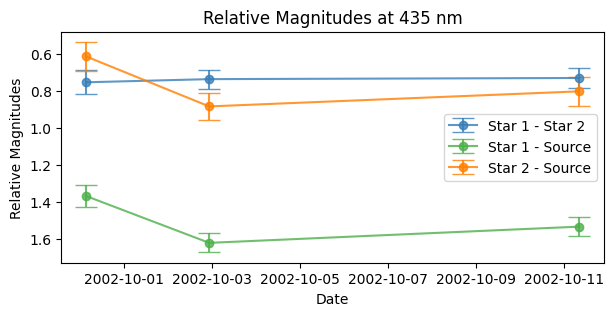}
\plotone{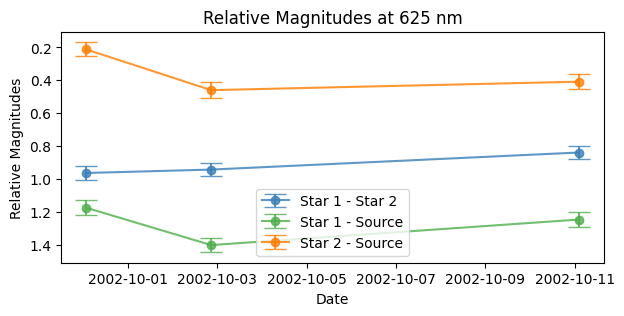}
\caption{Relative magnitude comparison between variable source and constant magnitude surrounding stars in filters 435 and 625 nm.}.
\label{fig:435}
\end{figure}

This variable was located at RA = 0:23:52.6444, Dec = -72:05:35.076. We calculated the change in relative magnitude of the variable source in comparison to two stars with constant magnitude. We calculated the standard deviation using  \texttt{ApertureStats} within \texttt{Photutils}. The change in magnitude of the variable exceeds the standard deviation in both 435 nm and 625 nm by approximately 5$\sigma$. 

\begin{figure}[t!]
\includegraphics[width=\linewidth]{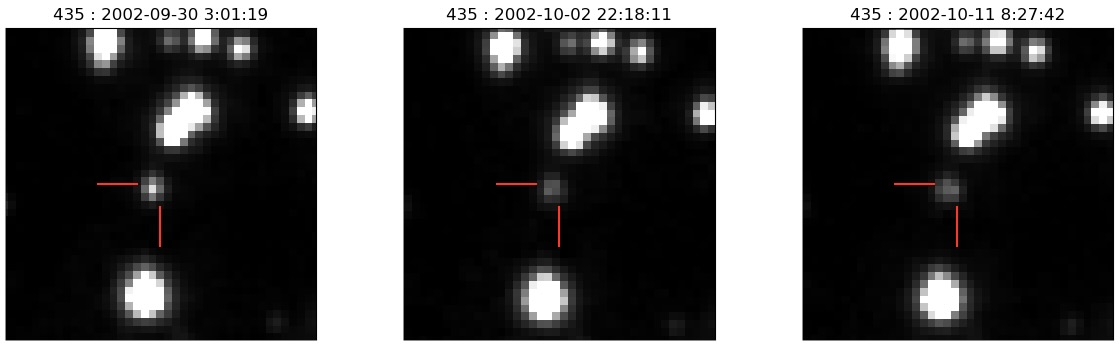}
\includegraphics[width=\linewidth]{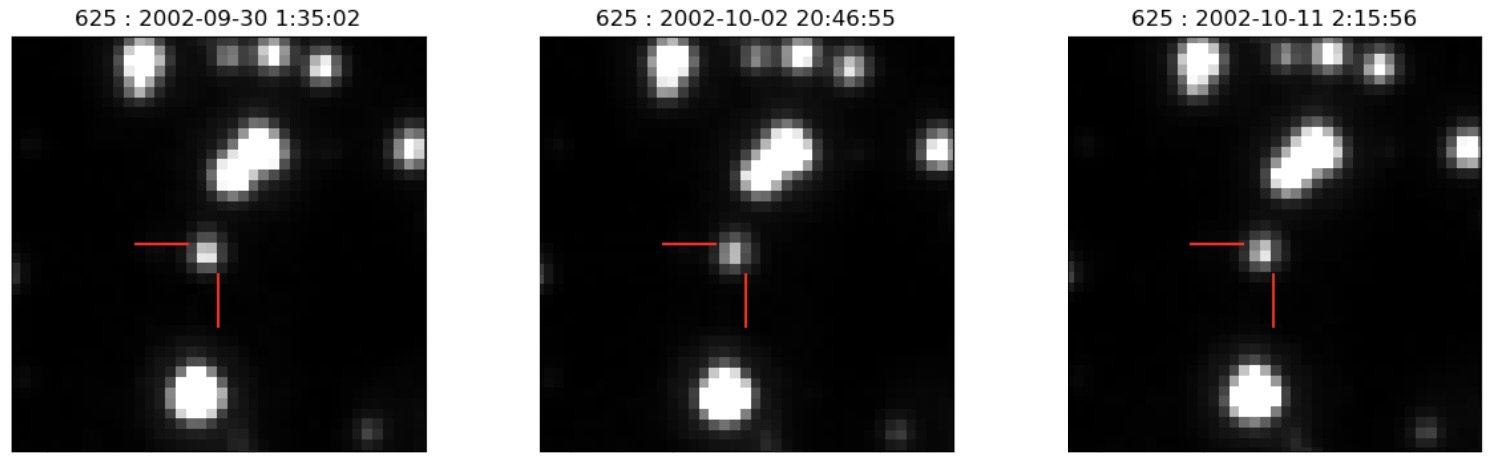}
\caption{Image comparison between variable object (centered) and constant magnitude surrounding stars in filters 435 nm (top) and 625 nm (bottom). The images shown correspond to the epochs the relative magnitudes shown in Figure \ref{fig:435} were calculated from.}
\label{fig:image_comparison}
\end{figure}

The results show a change of $\approx$ 0.3 Mag in 425 nm, and $\approx$ 0.2 Mag in 635nm between the 09-30-2002 and 10-11-2002. This change in magnitude in these observations does not meet the threshold of a dwarf nova eruption. Relative magnitude data are shown in Tables \ref{table:relmag435} and \ref{table:relmag625}. A plot of the light curve in relative magnitudes is shown in Figure \ref{fig:435}. A finding chart is visible in Figure \ref{fig:image_comparison}.

\section{Analysis}
The study functions as an exploration of a small subset of \textit{HST} data. While no objects were found with the sufficient change in magnitude to suggest a dwarf nova eruption, the small sample size of this study limits our ability to make larger conclusions about the CV population in 47 Tucanae. Though this study is not conclusive, our result should be considered in a larger study exploring other remaining unsearched data of 47 Tucanae.

\begin{figure}
    \centering
    \includegraphics[width=\linewidth]{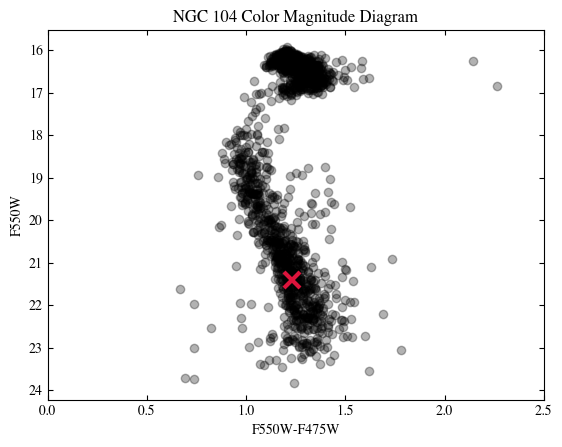}
    \caption{Color magnitude diagram of NGC 104. The unclassified variable source found in this study is denoted with a red 'X'. It sits on the main sequence, ruling out the possibility of its classification as a CV.}
    \label{fig:cmd}
\end{figure}

To eliminate the possibility that the variable was a DN near the end of its eruption, we placed it on a color magnitude diagram (CMD) for NGC 104. For the CMD, we selected datasets J6LL01010 and J6LL01YGQ (see Table \ref{epochs}) from the \textit{HST MAST} archive,  taken at F475W and F550W on the same date. We used \texttt{SEP} \citep{Barbary;2016}, a simplified Python library based on the original \texttt{Source Extractor} pipeline \citep{Bertin;1996}, to gather coordinates of stars in the cluster for aperture photometry. The resulting CMD is shown in Figure \ref{fig:cmd}, with the variable source denoted with an `X'. The source's placement on the cluster's main sequence rules out the possibility of it being a CV. A larger sample of time domain data of this object could be used to understand the cause for the variability, but that is beyond the scope of this project.

\begin{table}
\centering
\begin{tabular}{c|c|c|c}
Date & S1-Source Mag & S2-Source Mag & S1-S2 Mag \\
\hline 

2002-09-30 & $1.37 \pm 0.06$ & $0.61 \pm 0.06$ & $0.75 \pm 0.08$  \\
2002-10-02 & $1.62 \pm 0.05$ & $0.89 \pm 0.05$ & $0.74 \pm 0.07$ \\
2002-10-11 & $1.53 \pm 0.06$ & $0.80 \pm 0.05$ & $0.73 \pm 0.08$  \\
\end{tabular}
\caption{Relative Magnitudes of Variable Source at 435nm}\label{table:relmag435}
\centering
\begin{tabular}{c|c|c|c}
Date & S1-Source Mag & S2-Source Mag & S1-S2 Mag \\
\hline 

2002-09-30 & $1.175 \pm 0.043$ & $0.211 \pm 0.044$ & $0.964 \pm 0.043$ \\
2002-10-02& $1.403 \pm 0.042$ & $0.460 \pm 0.044$ & $0.944 \pm 0.047$ \\
2002-10-11& $1.248 \pm 0.042$ & $0.408 \pm 0.046$ & $0.839 \pm 0.047$ \\
\end{tabular}
\caption{Relative Magnitudes of Variable Source at 625nm}\label{table:relmag625}
Relative magnitudes of variable source and surrounding stars at 435nm and 625nm. ``S1'' and ``S2'' refer to surrounding stars of constant magnitude, and ``Source'' refers to the variable source.
\end{table}

% % \input{relmags425}
% \begin{table}\label{table:relmag625}
% \centering
% \begin{tabular}{c|c|c|c}
% Date & S1-Source Mag & S2-Source Mag & S1-S2 Mag \\
% \hline
% 2002-09-30 & 1.3846 & 0.6149 & 0.7697  \\
% 2002-10-02 & 1.6220&0.8850&0.7370 \\
% 2002-10-11 & 1.5437&0.80366&0.7401  \\
% \end{tabular}
% \caption{Relative Magnitudes of Variable Source at 425nm}
% \centering
% \begin{tabular}{c|c|c|c}
% Date & S1-Source Mag & S2-Source Mag & S1-S2 Mag \\
% \hline
% 2002-09-30 & 1.1950 &	0.3343 & 0.8607 \\
% 2002-10-02&1.4428 & 0.5818 & 0.8610 \\
% 2002-10-11&1.2998 & 0.5590 & 0.7408 \\
% \end{tabular}
% \caption{Relative Magnitudes of Variable Source at 635nm}
% Relative magnitudes of variable source and surrounding stars at 425nm and 625nm. ``S1'' and ``S2'' refer to surrounding stars of constant magnitude, and ``Source'' refers to the variable source. \label{table:relmag435}
% \end{table}

\vspace{20mm}

\section{Conclusion}
This study did not confirm any new DN or CV detections in 47 Tuc. One variable source was found, but the small change in magnitude of this source and main sequence CMD placement rule out the possibility of it being a dwarf nova. Further flux monitoring could identify the cause of variability.

This null result supports recent literature suggesting a different approach to CV detection. This could include different detection methodologies (searching in quiescence, FUV observations, H$\alpha$ emission) or revisiting predictions for detectable CV in GCs.

\section{Acknowledgements}
We would like to thank Dr. Eilat Glikman and Dr. Paul Hess at the Department of Physics at Middlebury College for facilitating this project. We would also like to acknowledge CUNY faculty Dr. Jillian Bellovary, Dr. Joshua Tan, and Dr. Jared Goldberg for their support in completing this work.

This research made use of Photutils, an Astropy package for
detection and photometry of astronomical sources \citep{bradley_2026_19636730}.

% \begin{figure*}
% \centering
% \includegraphics[width=0.3\linewidth]{findingchart.jpeg}
% \caption{Finding chart for the variable source found at RA = 0:23:52.6444, Dec. = -72:05:35.076. This image was taken on 09/30/02 at 625nm.}
% \label{finding}
% \end{figure*}

\vspace{5mm}

\facilities{\textit{HUBBLE}}
\software{Astropy \citep{2018AJ....156..123A}, Photutils \citep{bradley_2026_19636730}, Astrodrizzle \citep{Hack;2012}}
\pagebreak




\bibliographystyle{aasjournal}
\bibliography{biblio}

@software{bradley_2026_19636730,
  author       = {Bradley, Larry and
                  Sipőcz, Brigitta M. and
                  Robitaille, T. P. and
                  Tollerud, E. J. and
                  Vinícius, Zé and
                  Deil, Christoph and
                  Barbary, Kyle and
                  Wilson, Tom J. and
                  Busko, Ivo and
                  Donath, Axel and
                  Günther, Hans Moritz and
                  Cara, Mihai and
                  Lim, P. L. and
                  Meßlinger, Sebastian and
                  Conseil, Simon and
                  Droettboom, Michael and
                  Bostroem, K. Azalee and
                  Bray, E. M. and
                  Bratholm, Lars Andersen and
                  Burnett, Zach and
                  Jamieson, William and
                  Ginsburg, Adam and
                  Taranu, Dan and
                  Barentsen, Geert and
                  Craig, Matthew W. and
                  Morris, Brett M. and
                  Perrin, Marshall and
                  Rathi, Shivangee},
  title        = {Photutils},
  month        = apr,
  year         = 2026,
  publisher    = {Zenodo},
  version      = {3.0.0},
  doi          = {10.5281/zenodo.19636730},
  url          = {https://doi.org/10.5281/zenodo.19636730},
  swhid        = {swh:1:dir:5ea90432cd987336a26af201c1b50343b12e2daa
                   ;origin=https://doi.org/10.5281/zenodo.596036;visi
                   t=swh:1:snp:25dd84d95353ec3813baa8785c248017da7192
                   33;anchor=swh:1:rel:253cbde27398f9aea9d7368f386661
                   8c00fe16f9;path=astropy-photutils-3322558
                  },
}

@ARTICLE{RiveraSandoval,
       author = {{Rivera Sandoval}, L.~E. and {van den Berg}, M. and {Heinke}, C.~O. and {Cohn}, H.~N. and {Lugger}, P.~M. and {Anderson}, J. and {Cool}, A.~M. and {Edmonds}, P.~D. and {Wijnands}, R. and {Ivanova}, N. and {Grindlay}, J.~E.},
        title = "{New cataclysmic variables and other exotic binaries in the globular cluster 47 Tucanae$^{*}$}",
      journal = {\mnras},
         year = 2018,
        month = apr,
       volume = {475},
       number = {4},
        pages = {4841-4867},
          doi = {10.1093/mnras/sty058},
archivePrefix = {arXiv},
       eprint = {1705.07100},
 primaryClass = {astro-ph.SR},
       adsurl = {https://ui.adsabs.harvard.edu/abs/2018MNRAS.475.4841R}
}

@ARTICLE{Fabian,
       author = {{Fabian}, A.~C. and {Pringle}, J.~E. and {Rees}, M.~J.},
        title = "{Tidal capture formation of binary systems and X-ray sources in globular clusters.}",
      journal = {\mnras},
         year = 1975,
        month = aug,
       volume = {172},
        pages = {15},
          doi = {10.1093/mnras/172.1.15P},
       adsurl = {https://ui.adsabs.harvard.edu/abs/1975MNRAS.172P..15F}
}

@ARTICLE{Michaely,
       author = {{Michaely}, Erez and {Shara}, Michael M.},
        title = "{White dwarf-main sequence star collisions from wide triples in the field}",
      journal = {\mnras},
         year = 2021,
        month = apr,
       volume = {502},
       number = {3},
        pages = {4540-4546},
          doi = {10.1093/mnras/stab339},
archivePrefix = {arXiv},
       eprint = {2012.00022},
 primaryClass = {astro-ph.HE},
       adsurl = {https://ui.adsabs.harvard.edu/abs/2021MNRAS.502.4540M}
}

@ARTICLE{Vandenberg,
       author = {{Vandenberg}, Don A. and {Bolte}, Michael and {Stetson}, Peter B.},
        title = "{The Age of the Galactic Globular Cluster System}",
      journal = {\araa},
         year = 1996,
        month = jan,
       volume = {34},
        pages = {461-510},
          doi = {10.1146/annurev.astro.34.1.461},
       adsurl = {https://ui.adsabs.harvard.edu/abs/1996ARA&A..34..461V}
}

@ARTICLE{Lasota,
       author = {{Lasota}, Jean-Pierre},
        title = "{The disc instability model of dwarf novae and low-mass X-ray binary transients}",
      journal = {\nar},
         year = 2001,
        month = jun,
       volume = {45},
       number = {7},
        pages = {449-508},
          doi = {10.1016/S1387-6473(01)00112-9},
archivePrefix = {arXiv},
       eprint = {astro-ph/0102072},
 primaryClass = {astro-ph},
       adsurl = {https://ui.adsabs.harvard.edu/abs/2001NewAR..45..449L}
}

@ARTICLE{2018AJ....156..123A,
       author = {{Astropy Collaboration} and {Price-Whelan}, A.~M. and {Sip{\H{o}}cz}, B.~M. and {G{\"u}nther}, H.~M. and {Lim}, P.~L. and {Crawford}, S.~M. and {Conseil}, S. and {Shupe}, D.~L. and {Craig}, M.~W. and {Dencheva}, N. and {Ginsburg}, A. and {VanderPlas}, J.~T. and {Bradley}, L.~D. and {P{\'e}rez-Su{\'a}rez}, D. and {de Val-Borro}, M. and {Aldcroft}, T.~L. and {Cruz}, K.~L. and {Robitaille}, T.~P. and {Tollerud}, E.~J. and {Ardelean}, C. and {Babej}, T. and {Bach}, Y.~P. and {Bachetti}, M. and {Bakanov}, A.~V. and {Bamford}, S.~P. and {Barentsen}, G. and {Barmby}, P. and {Baumbach}, A. and {Berry}, K.~L. and {Biscani}, F. and {Boquien}, M. and {Bostroem}, K.~A. and {Bouma}, L.~G. and {Brammer}, G.~B. and {Bray}, E.~M. and {Breytenbach}, H. and {Buddelmeijer}, H. and {Burke}, D.~J. and {Calderone}, G. and {Cano Rodr{\'\i}guez}, J.~L. and {Cara}, M. and {Cardoso}, J.~V.~M. and {Cheedella}, S. and {Copin}, Y. and {Corrales}, L. and {Crichton}, D. and {D'Avella}, D. and {Deil}, C. and {Depagne}, {\'E}. and {Dietrich}, J.~P. and {Donath}, A. and {Droettboom}, M. and {Earl}, N. and {Erben}, T. and {Fabbro}, S. and {Ferreira}, L.~A. and {Finethy}, T. and {Fox}, R.~T. and {Garrison}, L.~H. and {Gibbons}, S.~L.~J. and {Goldstein}, D.~A. and {Gommers}, R. and {Greco}, J.~P. and {Greenfield}, P. and {Groener}, A.~M. and {Grollier}, F. and {Hagen}, A. and {Hirst}, P. and {Homeier}, D. and {Horton}, A.~J. and {Hosseinzadeh}, G. and {Hu}, L. and {Hunkeler}, J.~S. and {Ivezi{\'c}}, {\v{Z}}. and {Jain}, A. and {Jenness}, T. and {Kanarek}, G. and {Kendrew}, S. and {Kern}, N.~S. and {Kerzendorf}, W.~E. and {Khvalko}, A. and {King}, J. and {Kirkby}, D. and {Kulkarni}, A.~M. and {Kumar}, A. and {Lee}, A. and {Lenz}, D. and {Littlefair}, S.~P. and {Ma}, Z. and {Macleod}, D.~M. and {Mastropietro}, M. and {McCully}, C. and {Montagnac}, S. and {Morris}, B.~M. and {Mueller}, M. and {Mumford}, S.~J. and {Muna}, D. and {Murphy}, N.~A. and {Nelson}, S. and {Nguyen}, G.~H. and {Ninan}, J.~P. and {N{\"o}the}, M. and {Ogaz}, S. and {Oh}, S. and {Parejko}, J.~K. and {Parley}, N. and {Pascual}, S. and {Patil}, R. and {Patil}, A.~A. and {Plunkett}, A.~L. and {Prochaska}, J.~X. and {Rastogi}, T. and {Reddy Janga}, V. and {Sabater}, J. and {Sakurikar}, P. and {Seifert}, M. and {Sherbert}, L.~E. and {Sherwood-Taylor}, H. and {Shih}, A.~Y. and {Sick}, J. and {Silbiger}, M.~T. and {Singanamalla}, S. and {Singer}, L.~P. and {Sladen}, P.~H. and {Sooley}, K.~A. and {Sornarajah}, S. and {Streicher}, O. and {Teuben}, P. and {Thomas}, S.~W. and {Tremblay}, G.~R. and {Turner}, J.~E.~H. and {Terr{\'o}n}, V. and {van Kerkwijk}, M.~H. and {de la Vega}, A. and {Watkins}, L.~L. and {Weaver}, B.~A. and {Whitmore}, J.~B. and {Woillez}, J. and {Zabalza}, V. and {Astropy Contributors}},
        title = "{The Astropy Project: Building an Open-science Project and Status of the v2.0 Core Package}",
      journal = {\aj},
         year = 2018,
        month = sep,
       volume = {156},
       number = {3},
          eid = {123},
        pages = {123},
          doi = {10.3847/1538-3881/aabc4f},
archivePrefix = {arXiv},
       eprint = {1801.02634},
 primaryClass = {astro-ph.IM},
       adsurl = {https://ui.adsabs.harvard.edu/abs/2018AJ....156..123A}
}

@ARTICLE{distef,
       author = {{Di Stefano}, R. and {Rappaport}, S.},
        title = "{Predictions of a Population of Cataclysmic Variables in Globular Clusters}",
      journal = {\apj},
         year = 1994,
        month = mar,
       volume = {423},
        pages = {274},
          doi = {10.1086/173805},
       adsurl = {https://ui.adsabs.harvard.edu/abs/1994ApJ...423..274D}
}

@ARTICLE{Harris,
       author = {{Harris}, William E.},
        title = "{A Catalog of Parameters for Globular Clusters in the Milky Way}",
      journal = {\aj},
         year = 1996,
        month = oct,
       volume = {112},
        pages = {1487},
          doi = {10.1086/118116},
       adsurl = {https://ui.adsabs.harvard.edu/abs/1996AJ....112.1487H}
}

@ARTICLE{robinson,
       author = {{Robinson}, E.~L.},
        title = "{The structure of cataclysmic variables.}",
      journal = {\araa},
         year = 1976,
        month = jan,
       volume = {14},
        pages = {119-142},
          doi = {10.1146/annurev.aa.14.090176.001003},
       adsurl = {https://ui.adsabs.harvard.edu/abs/1976ARA&A..14..119R}
}

@ARTICLE{sharaberg,
       author = {{Shara}, Michael M. and {Bergeron}, Louis E. and {Gilliland}, Ronald L. and {Saha}, Abhijit and {Petro}, Larry},
        title = "{Cataclysmic and Close Binaries in Star Clusters. IV. The Unexpectedly Low Number of Erupting Dwarf Novae Detected by the Hubble Space Telescope in the Core of 47 Tucanae}",
      journal = {\apj},
         year = 1996,
        month = nov,
       volume = {471},
        pages = {804},
          doi = {10.1086/178008},
       adsurl = {https://ui.adsabs.harvard.edu/abs/1996ApJ...471..804S}
}

@BOOK{Carroll;1996,
       author = {{Carroll}, Bradley W. and {Ostlie}, Dale A.},
        title = "{An Introduction to Modern Astrophysics}",
        publisher = "Pearson",
         year = 1996,
       adsurl = {https://ui.adsabs.harvard.edu/abs/1996ima..book.....C}
}

@ARTICLE{Shara;2005,
       author = {{Shara}, Michael M. and {Hinkley}, Sasha and {Zurek}, David R.},
        title = "{Cataclysmic and Close Binaries in Star Clusters. V. Erupting Dwarf Novae, Faint Blue Stars, X-Ray Sources, and the Classical Nova in the Core of M80}",
      journal = {\apj},
         year = 2005,
        month = dec,
       volume = {634},
       number = {2},
        pages = {1272-1285},
          doi = {10.1086/497101},
archivePrefix = {arXiv},
       eprint = {astro-ph/0508406},
 primaryClass = {astro-ph},
       adsurl = {https://ui.adsabs.harvard.edu/abs/2005ApJ...634.1272S}
}

@ARTICLE{Servillat;2011,
       author = {{Servillat}, M. and {Webb}, N.~A. and {Lewis}, F. and {Knigge}, C. and {van den Berg}, M. and {Dieball}, A. and {Grindlay}, J.},
        title = "{A Dwarf Nova in the Globular Cluster M13}",
      journal = {\apj},
         year = 2011,
        month = jun,
       volume = {733},
       number = {2},
          eid = {106},
        pages = {106},
          doi = {10.1088/0004-637X/733/2/106},
archivePrefix = {arXiv},
       eprint = {1103.4638},
 primaryClass = {astro-ph.SR},
       adsurl = {https://ui.adsabs.harvard.edu/abs/2011ApJ...733..106S}
}

@ARTICLE{Knigge;2003,
       author = {{Knigge}, Christian and {Zurek}, David R. and {Shara}, Michael M. and {Long}, Knox S. and {Gilliland}, Ronald L.},
        title = "{A Far-Ultraviolet Survey of 47 Tucanae. II. The Long-Period Cataclysmic Variable AKO 9}",
      journal = {\apj},
         year = 2003,
        month = dec,
       volume = {599},
       number = {2},
        pages = {1320-1332},
          doi = {10.1086/379609},
archivePrefix = {arXiv},
       eprint = {astro-ph/0309168},
 primaryClass = {astro-ph},
       adsurl = {https://ui.adsabs.harvard.edu/abs/2003ApJ...599.1320K}
}

@ARTICLE{Belloni;2016,
       author = {{Belloni}, Diogo and {Giersz}, Mirek and {Askar}, Abbas and {Leigh}, Nathan and {Hypki}, Arkadiusz},
        title = "{MOCCA-SURVEY database I. Accreting white dwarf binary systems in globular clusters - I. Cataclysmic variables - present-day population}",
      journal = {\mnras},
         year = 2016,
        month = nov,
       volume = {462},
       number = {3},
        pages = {2950-2969},
          doi = {10.1093/mnras/stw1841},
archivePrefix = {arXiv},
       eprint = {1607.07619},
 primaryClass = {astro-ph.GA},
       adsurl = {https://ui.adsabs.harvard.edu/abs/2016MNRAS.462.2950B}
}

@ARTICLE{Cool;1995,
       author = {{Cool}, Adrienne M. and {Grindlay}, Jonathan E. and {Bailyn}, Charles D. and {Callanan}, Paul J. and {Hertz}, Paul},
        title = "{The Faint X-Ray Sources In and Out of omega Centauri: X-Ray Observations and Optical Identifications}",
      journal = {\apj},
         year = 1995,
        month = jan,
       volume = {438},
        pages = {719},
          doi = {10.1086/175116},
       adsurl = {https://ui.adsabs.harvard.edu/abs/1995ApJ...438..719C}
}

@ARTICLE{Grindlay;1995,
       author = {{Grindlay}, J.~E. and {Cool}, A.~M. and {Callanan}, P.~J. and {Bailyn}, C.~D. and {Cohn}, H.~N. and {Lugger}, P.~M.},
        title = "{Spectroscopic Identification of Probable Cataclysmic Variables in the Globular Cluster NGC 6397}",
      journal = {\apjl},
         year = 1995,
        month = dec,
       volume = {455},
        pages = {L47},
          doi = {10.1086/309806},
       adsurl = {https://ui.adsabs.harvard.edu/abs/1995ApJ...455L..47G}
}

@ARTICLE{Pietrukowicz;2009,
       author = {{Pietrukowicz}, P.},
        title = "{H{\ensuremath{\alpha}} Imaging of X-ray Sources in Selected Globular Clusters with the SOAR Telescope}",
      journal = {\actaa},
         year = 2009,
        month = sep,
       volume = {59},
       number = {3},
        pages = {291-305},
          doi = {10.48550/arXiv.0910.1302},
archivePrefix = {arXiv},
       eprint = {0910.1302},
 primaryClass = {astro-ph.GA},
       adsurl = {https://ui.adsabs.harvard.edu/abs/2009AcA....59..291P}
}

@ARTICLE{Dieball;2010,
       author = {{Dieball}, Andrea and {Long}, Knox S. and {Knigge}, Christian and {Thomson}, Grace S. and {Zurek}, David R.},
        title = "{A Far-ultraviolet Survey of M80: X-Ray Source Counterparts, Strange Blue Stragglers, and the Recovery of Nova T Sco}",
      journal = {\apj},
         year = 2010,
        month = feb,
       volume = {710},
       number = {1},
        pages = {332-345},
          doi = {10.1088/0004-637X/710/1/332},
archivePrefix = {arXiv},
       eprint = {1001.1161},
 primaryClass = {astro-ph.SR},
       adsurl = {https://ui.adsabs.harvard.edu/abs/2010ApJ...710..332D}
}

@ARTICLE{Stodolkiewicz;1986,
       author = {{Stodolkiewicz}, J.~S.},
        title = "{Dynamical evolution of globular clusters. II - Binaries Method}",
      journal = {\actaa},
         year = 1986,
        month = jan,
       volume = {36},
       number = {1},
        pages = {19-41},
       adsurl = {https://ui.adsabs.harvard.edu/abs/1986AcA....36...19S}
}

@ARTICLE{Hut;1992,
       author = {{Hut}, Piet and {McMillan}, Steve and {Goodman}, Jeremy and {Mateo}, Mario and {Phinney}, E.~S. and {Pryor}, Carlton and {Richer}, Harvey B. and {Verbunt}, Frank and {Weinberg}, Martin},
        title = "{Binaries in Globular Clusters}",
      journal = {\pasp},
         year = 1992,
        month = nov,
       volume = {104},
        pages = {981},
          doi = {10.1086/133085},
       adsurl = {https://ui.adsabs.harvard.edu/abs/1992PASP..104..981H}
}

@ARTICLE{RiveraSandoval;2024,
       author = {{Rivera Sandoval}, Liliana and {Belloni}, Diogo and {Ramos Arevalo}, Miriam},
        title = "{Reevaluating the Origin of Detectable Cataclysmic Variables in Globular Clusters: Testing the Importance of Dynamics}",
      journal = {\apjl},
         year = 2024,
        month = apr,
       volume = {964},
       number = {2},
          eid = {L20},
        pages = {L20},
          doi = {10.3847/2041-8213/ad2c8f},
archivePrefix = {arXiv},
       eprint = {2404.03796},
 primaryClass = {astro-ph.SR},
       adsurl = {https://ui.adsabs.harvard.edu/abs/2024ApJ...964L..20R}
}

@ARTICLE{Baumgardt2018,
       author = {{Baumgardt}, H. and {Hilker}, M.},
        title = "{A catalogue of masses, structural parameters, and velocity dispersion profiles of 112 Milky Way globular clusters}",
      journal = {\mnras},
         year = 2018,
        month = aug,
       volume = {478},
       number = {2},
        pages = {1520-1557},
          doi = {10.1093/mnras/sty1057},
archivePrefix = {arXiv},
       eprint = {1804.08359},
 primaryClass = {astro-ph.GA},
       adsurl = {https://ui.adsabs.harvard.edu/abs/2018MNRAS.478.1520B}
}

@INPROCEEDINGS{Belloni2020,
       author = {{Belloni}, Diogo and {Giersz}, Mirek and {Sandoval}, Liliana E. Rivera and {Askar}, Abbas and {Ciecielag}, Pawel},
        title = "{Are most Cataclysmic Variables in Globular Clusters dynamically formed?}",
    booktitle = {Star Clusters: From the Milky Way to the Early Universe},
         year = 2020,
       editor = {{Bragaglia}, Angela and {Davies}, Melvyn and {Sills}, Alison and {Vesperini}, Enrico},
       series = {IAU Symposium},
       volume = {351},
        month = jan,
        pages = {404-407},
          doi = {10.1017/S174392131900718X},
archivePrefix = {arXiv},
       eprint = {1907.06527},
 primaryClass = {astro-ph.GA},
       adsurl = {https://ui.adsabs.harvard.edu/abs/2020IAUS..351..404B}
}

@ARTICLE{Bertin;1996,
       author = {{Bertin}, E. and {Arnouts}, S.},
        title = "{SExtractor: Software for source extraction.}",
      journal = {\aaps},
         year = 1996,
        month = jun,
       volume = {117},
        pages = {393-404},
          doi = {10.1051/aas:1996164},
       adsurl = {https://ui.adsabs.harvard.edu/abs/1996A&AS..117..393B}
}

@ARTICLE{Barbary;2016,
       author = {{Barbary}, Kyle},
        title = "{SEP: Source Extractor as a library}",
      journal = {The Journal of Open Source Software},
         year = 2016,
        month = oct,
       volume = {1},
       number = {6},
          eid = {58},
        pages = {58},
          doi = {10.21105/joss.00058},
       adsurl = {https://ui.adsabs.harvard.edu/abs/2016JOSS....1...58B}
}

@INPROCEEDINGS{Hack;2012,
       author = {{Hack}, Warren J. and {Dencheva}, N. and {Fruchter}, A.~S. and {Armstrong}, A. and {Avila}, R. and {Baggett}, S. and {Bray}, E. and {Droettboom}, M. and {Dulude}, M. and {Gonzaga}, S. and {Grogin}, N.~A. and {Kozhurina-Platais}, V. and {Lucas}, R.~A. and {Mack}, J. and {MacKenty}, J. and {Petro}, L. and {Pirzkal}, N. and {Rajan}, A. and {Smith}, L.~J. and {Sontag}, C. and {Ubeda}, L.},
        title = "{AstroDrizzle: More than a New MultiDrizzle}",
    booktitle = {American Astronomical Society Meeting Abstracts \#220},
         year = 2012,
       series = {American Astronomical Society Meeting Abstracts},
       volume = {220},
        month = may,
          eid = {135.15},
        pages = {135.15},
       adsurl = {https://ui.adsabs.harvard.edu/abs/2012AAS...22013515H}
}



\end{document}